\documentclass[twocolumn,aps,prb,showpacs,superscriptaddress]{revtex4-1}
\pdfoutput=1
\usepackage{amssymb}
\usepackage{graphicx}
\usepackage[pdftex]{hyperref}
\hypersetup{colorlinks=true,citecolor=blue,linkcolor=blue,urlcolor=blue}
\usepackage[all]{hypcap}
\usepackage{epstopdf}
\usepackage{bm}      
\usepackage{rotating}
\usepackage{color}

\begin{document}

\title{Energetics and carrier transport in doped Si/SiO$_2$ quantum dots}

\author{Nuria Garcia-Castello}
\email{nuriagarcia@ub.edu}
\affiliation{MIND-IN$^2$UB, Department of Electronics, Universitat de Barcelona, C/Mart\'i i Franqu\`es 1, E-08028 Barcelona, Spain.}

\author{Sergio Illera}
\affiliation{MIND-IN$^2$UB, Department of Electronics, Universitat de Barcelona, C/Mart\'i i Franqu\`es 1, E-08028 Barcelona, Spain.}

\author{Joan Daniel Prades}
\affiliation{MIND-IN$^2$UB, Department of Electronics, Universitat de Barcelona, C/Mart\'i i Franqu\`es 1, E-08028 Barcelona, Spain.}

\author{Stefano Ossicini}
\affiliation{Dipartimento di Scienze e Metodi dell'Ingegneria and Centro Interdipartimentale En\&Tech, Universit\`a di Modena e Reggio Emilia, via Amendola 2 Pad. Morselli, I-42122 Reggio Emilia, Italy.}

\author{Albert Cirera}
\affiliation{MIND-IN$^2$UB, Department of Electronics, Universitat de Barcelona, C/Mart\'i i Franqu\`es 1, E-08028 Barcelona, Spain.}

\author{Roberto Guerra}
\email{guerra@sissa.it}
\affiliation{International School for Advanced Studies (SISSA), Via Bonomea 265, I-34136 Trieste, Italy; CNR-IOM Democritos National Simulation Center,Via Bonomea 265, I-34136 Trieste, Italy.}

\begin{abstract}
In the present theoretical work we have considered impurities, either boron or phosphorous, located at different substitutional sites in silicon quantum dots (Si-QDs) with diameters around 1.5\,nm, embedded in a SiO$_2$ matrix. Formation energy calculations reveal that the most energetically-favored doping sites are inside the QD and at the Si/SiO$_2$ interface for P and B impurities, respectively. Furthermore, electron and hole transport calculations show in all the cases a strong reduction of the minimum voltage threshold, and a corresponding increase of the total current in the low-voltage regime. At higher voltage, our findings indicate a significant increase of transport only for P-doped Si-QDs, while the electrical response of B-doped ones does not stray from the undoped case. These findings are of support for the employment of doped Si-QDs in a wide range of applications, such as Si-based photonics or photovoltaic solar cells.

\noindent[ Electronic Supplementary Information (ESI) available. See DOI: 10.1039/c5nr02616d ]
\end{abstract}

\maketitle

\section{Introduction}\label{sec.intro}

Semiconductor quantum dots (QDs) are promising structures due to their tunable band gap with QD diameter. Silicon QDs (Si-QDs) are, among all, the ideal candidates for mass-scale devices production, because of the abundancy of silicon and its non-toxic, bio-compatible, and ecologic nature. Embedding Si-QDs in a dielectric matrix is one way to obtain an efficient quantum-confinement effect.\cite{QC} Besides, the possibility of introducing dopant atoms has been suggested to improve the the achievable macroscopic currents in QD-based devices.

Doping of Si-QDs embedded in silica has been already investigated by several experimental works.\cite{conibeer-2012,exp1,exp2,exp3,di,conibeer-2013,hao-B,hao-P,xie-B,perego,cho-P,B-surface,park-P,simonds-P}
In particular, it has been shown that electrically-activated impurity atoms located in substitutional sites tend to enhance the conductivity.\cite{di,conibeer-2013,hao-B,hao-P,xie-B}
Theoretically, several works have studied the formation and ionization energies, and the opto-electronic properties of freestanding doped Si-QDs.\cite{melnikov,ossicini-2006,mavros,eom,ma,location,zhou,ramos,mavros-54,carvalho1,carvalho2}
Instead, only recently theoretical works dealing with structural properties of doped embedded Si-QDs have appeared in literature.\cite{guerrajacs}

In any case, all the above works show that the final properties of these systems are strongly sensitive to the concentration and position of the impurities.
This fact makes necessary the accurate control of the impurities at the nanoscale in order to ensure repeatability.

Thanks to the recent advances, it is nowadays possible to dope Si-QDs with few\cite{weber} or even only one\cite{fuechsle} dopant atoms, and to experimentally obtain the density of states of the single QD.\cite{wolf}
With these premises, a comprehensive understanding of structural, electrical and transport properties of doped Si-QDs is hopefully going to be achieved soon. The aim of the present work is to shed light in this direction.
Theoretical simulations can provide a strong support in understanding the role of impurities in nanostructures, thanks to the possibility of manipulating the samples at the atomic level, and to the recent advance in the computing capabilities.

Here we report a theoretical study of electron and hole transport induced by B or P substitutional doping in a crystalline Si-QD embedded in SiO$_2$, for three different QDs. The structures with the lowest formation energies are identified, and the $I$-$V$ characteristic is obtained by a novel approach (See Method).
To our knowledge, this is the first study reporting theoretical $I$-$V$ characteristics of doped embedded Si-QDs.

\section{Structures and Method}\label{sec.method}

Despite the tremendous progress in the computational power made with the advent of supercomputers, a complete theoretical description of transport in large nanostructures is still far to be achieved. Approximations must be adopted in order to limit the computational effort, like using a reduced system size, or employing a simplified approach.
Most of the available studies on single- and two-QDs have been performed by using non-equilibrium Green functions formalism (NEGFF) with constant transition rates between QDs and one energy level per QD\cite{NDR,NEGF1,NEGF2,NEGF3,transport1,transport2}, and by using tunneling transmission coefficients with planar Si/SiO$_2$ values for the barrier height, and bulk-Si band gap.\cite{carreras,conibeer,solar2,transport3}

Here we make use of a different approach,\cite{sergio0,sergio1,sergio2} based on the transfer Hamiltonian formalism and non-coherent rate equations to describe the current, that takes into account the local potential due to the QD charge, computed in a self-consistently field regime with the non-equilibrium distribution function of the QD, and able to use more than one energy state per QD. In particular, we can use the density of states computed by {\it ab initio} calculations, a difficult issue to treat with NEGFF, allowing us to include implicitly the effect of dopant atoms in the transport properties.
With the same approach we investigated, in a previous work, the influence of QD size and amorphization level on the transport properties of undoped Si-QDs.\cite{IV-diameters}

\begin{figure}[t!]
	\centering
	\includegraphics[width=\columnwidth]{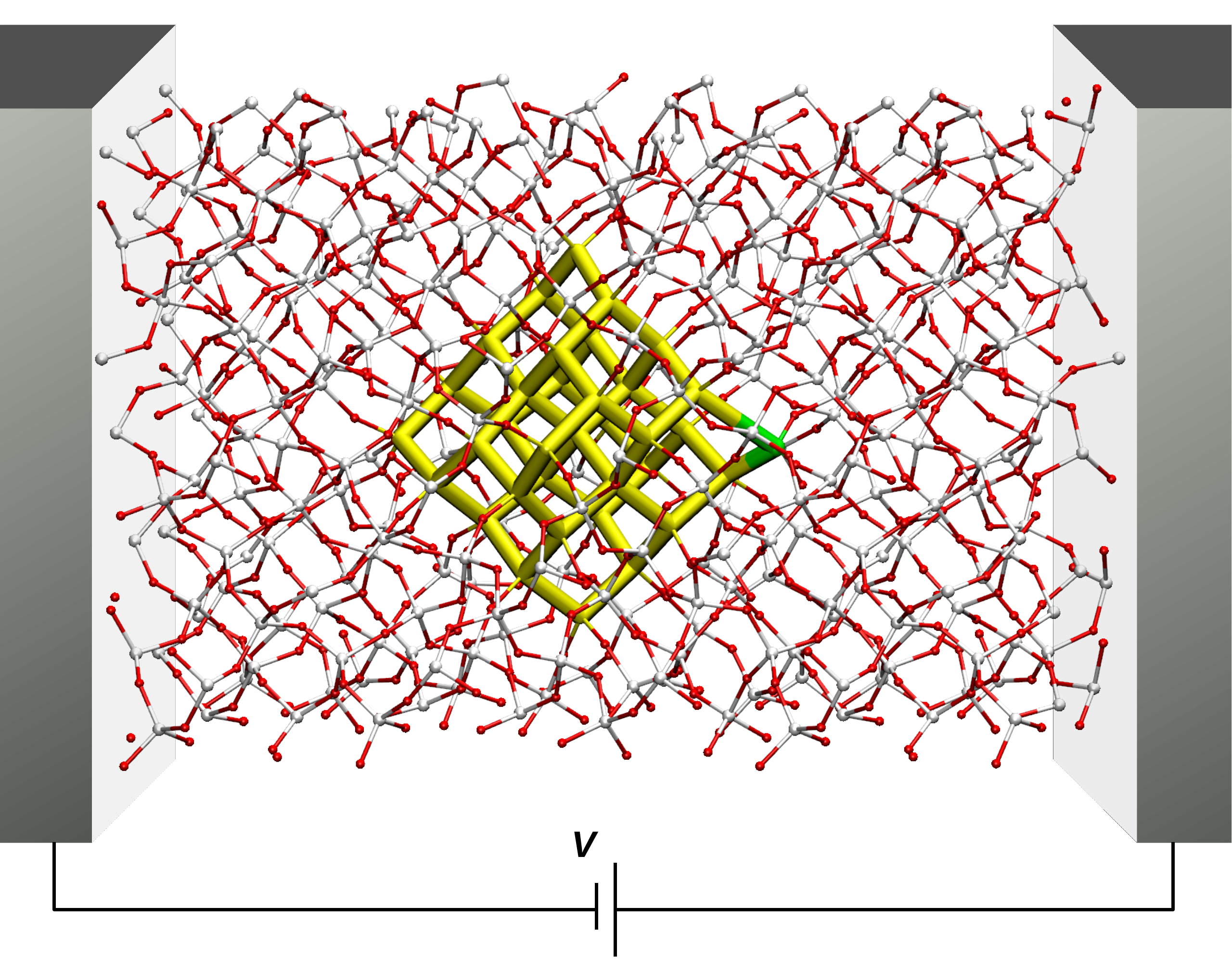}
	\caption{Sticks-and-balls representation of a Si$_{35}$ QD embedded in SiO$_2$ ($\sim$600 atoms), doped at the interface, and enclosed between two semi-infinite metallic leads with an applied voltage.}
	\label{fig:system}
\end{figure}

We compute the $I$-$V$ characteristic between two metallic semi-infinite electrodes coupled to an elastic scattering region (corresponding to the doped Si-QD embedded in the silica matrix) when an external bias voltage $V$ is applied (see Fig.~\ref{fig:system}). The expression of the current under the transfer Hamiltonian formalism is\cite{TF1,TF2}
\begin{equation}
	I=\frac{4\pi q}{\hbar} \displaystyle\int \frac{T_LT_R\rho_L\rho_{QD}\rho_R}{T_L\rho_L+T_R\rho_R}(f_L-f_R) \ dE~,
	\label{eq:IV}
\end{equation}
where $T_{L,R}(E)$ are the transmission probabilities between the left lead and the QD, and between the QD and the right lead, respectively; $\rho_{L,R,QD}(E)$ are the density of states of each region of the system, and $f_{L,R}(E)$ are the Fermi-Dirac distribution functions of the electrodes.

All the calculations are done at room temperature ($k_B$$T$ = 0.026 eV), and $\rho_{L/R}$ are assumed constant in energy. In principle, the presence of nanostructured contact could be described in our model making use of calculated $\rho_L(E)$ and $\rho_R(E)$ from atomistic leads, like e.g.\ gold tips. However, as indicated by previous works,\cite{ventra} in the latter case we expect no major change of the here-presented I-V curves, but rather a reduction of the current magnitude depending on the tips DOS. Clearly, for very small (molecular-like) electrodes$+$QD systems, currents become more sensitive to the geometrical conditions and a full {\it ab-initio} approach is required in that case.\cite{dai_zeng}

The transmission probabilities are calculated using WKB approximation of Fowler-Nordheim and direct tunnel mechanisms, which are the two more relevant tunneling mechanisms in QDs inside dielectric matrices.\cite{transmission} We set the distance between the Si-QD and each lead to 1.1\,nm for all the systems, the relative dielectric constant of the oxide to 3.9, and the oxide effective mass of electrons and holes to 0.40\,${m_e}$ and 0.32\,${m_e}$, respectively,\cite{carreras} ${m_e}$ being the free-electron mass.

The effect of charge inside the QD induced by the applied $V$ is taken into account. Thus, we solve self-consistently the equations for the local potential inside the QD, and the QD non-equilibrium distribution function. The details of this method are reported elsewhere.\cite{sergio0,sergio1,sergio2}
For the present study we assume the same capacitive coupling between the QD and the leads, yielding specular current trends for negative $V$.\cite{sergio1} Thus, to avoid redundancy we report only currents for positive applied $V$.

Assuming ballistic transport we have independent conduction channels for electrons and holes. The current for each carrier type can be calculated from Eq.~\ref{eq:IV} using the corresponding transmission coefficient and density of states. The total current is given by the sum of electron and hole currents.

The density of states of the Si-QD $\rho_{QD}$ has been computed within density functional theory (DFT) using the {\small SIESTA} code\cite{siesta1,siesta2}, and gaussian broadening of 0.05\,eV. It corresponds to the PDOS of the Si atoms of the QD together with the interface O atoms, in order to include the interface states.\cite{DFT1} The embedded Si-QDs of 32, 35, and 47 Si atoms (i.e. Si$_{32}$, Si$_{35}$, and Si$_{47}$) were obtained from a 3$\times$3$\times$3 $\beta$-cristobalite supercell, Si$_{216}$O$_{432}$ of side of 21.48\,\AA, by removing all the O atoms inside a cut-off spheres of given radius. In this way, no dangling bonds or defects are present, and all the O atoms at the interface are single-bonded to the Si atoms of the QD. The so-obtained embedded systems are formed by a total of about 600 atoms.
Spin polarized calculations were performed using norm-conserving Troullier-Martins\cite{pseudo} pseudopotentials with nonlinear core corrections within the local density approximation (LDA), with a Ceperley-Alder\cite{LDA1} exchange-correlation potential, as parameterized by Perdew-Zunger.\cite{LDA2} A cut-off of 250\,Ry on the electron density and no additional external pressure or stress were applied. All the calculations were performed at the $\Gamma$-point of reciprocal space, with an electronic temperature of 300\,K, and standard double-$\zeta$ basis set for all the atoms. Atomic positions and cell parameters were left totally free to move, with a force threshold of 0.01\,eV/\AA.

Thanks to the metastable nature of $\beta$-cristobalite, after relaxation all the SiO$_2$ in the supercell gets amorphized due to the presence of the QD. Structural and optical properties of the embedding SiO$_2$ match well those of a ``true'' amorphous SiO$_2$ glass (a-SiO$_2$), formed by annealing.\cite{DFT1}

As reported in a previous study\cite{IV-diameters} the presence of quantum confinement makes valence band offset (VBO) and conduction band offset (CBO) between Si-QDs and SiO$_2$ significantly different than in bulk or planar systems. In order to evaluate the band offset between SiO$_2$ and QD, we have aligned the DOS of an a-SiO$_2$ sample with that of the embedded Si-QD by matching the strong deep-valence peak of a-SiO$_2$, which is well observable in all the considered structures. Thus, we have obtained the VBO as the difference between QD and SiO$_2$ HOMOs (highest occupied molecular orbitals), and the CBO as the difference between SiO$_2$ and QD LUMOs (lowest unoccupied molecular orbitals). We also have defined the hole barrier (HB) as the difference between the Fermi energy ($E_F$) and the HOMO of the embedding a-SiO$_2$, and the electron barrier (EB) as the difference between the LUMO of the embedding a-SiO$_2$ and $E_F$. Since $E_F$ is approximately located at mid-$E_g$, it is HB\,$=$\,VBO+$E_g$/2, and EB\,=\,CBO+$E_g$/2.

The computed DFT HOMO-LUMO gap $E_g$ for a-SiO$_2$ and bulk-Si are 7.0\,eV and 0.6\,eV, respectively, in agreement with other calculations,\cite{LDA-LCAO} and smaller than the experimental values of 9.0\,eV and 1.1\,eV, respectively.
It is well known that Kohn-Sham eigenvalues give an underestimated $E_g$ due to the use of approximated exchange-correlation functionals. A correction to the fundamental band gap can be obtained by many-body calculations accounting for quasiparticle energies and excitonic corrections.\cite{DFT1} However, while the total correction to $E_g$ is noticeable in bulk materials, in strongly confined systems the enhanced excitonic interaction is known to compensate the self-energy of about the same amount.\cite{ramos,DFT1,osso1,osso2} As a consequence, in the case of small embedded QD, one deals with ``correct'' $E_g$ values (determined by QD states), but ``uncorrect'' band offsets due to the systematic error in the SiO$_2$-related energy values.

In the case of a Si/SiO$_2$ slab calculation in the bulk limit, we have obtained VBO and CBO of 2.6\,eV and 3.9\,eV, respectively, to be compared with the experimental values of 4.6\,eV (VBO) and 3.1\,eV (CBO).\cite{SiQD1,conibeer} Therefore, to compensate such deviations, we have applied a correction of 2.0\,eV to VBO and HB values, and of -0.8\,eV to CBO and EB values, while leaving $E_g$ unchanged. Since our QD size range is small, we apply the same correction for all the samples. Note that our uncorrected band offset match that of other works investigating charge-carrier transport in Si-QDs by hopping mechanisms.\cite{DFT7,Seino1}

We have positioned the impurity atom in three different substitutional sites in the embedded system: at the QD center (in the following ``dot''), at the QD/SiO$_2$ interface, and in the SiO$_2$ far away from the QD (in the following ``silica''). The Si atoms at the interface can form three possible suboxide types, Si$^{1+}$, Si$^{2+}$, Si$^{3+}$, depending on the number of bonded O atoms (in the following ``int-1'', ``int-2'', and ``int-3''). While Si$_{47}$ QD presents all the three suboxide types, Si$_{32}$ presents only Si$^{1+}$ and Si$^{3+}$ types, while Si$_{35}$ presents only Si$^{1+}$ and Si$^{2+}$ types.

\section{Results and Discussion}\label{sec.results}

\subsection{Structural properties}\label{sec.structures}

\begin{figure}[b!]
	\centering
	\includegraphics[width=\columnwidth]{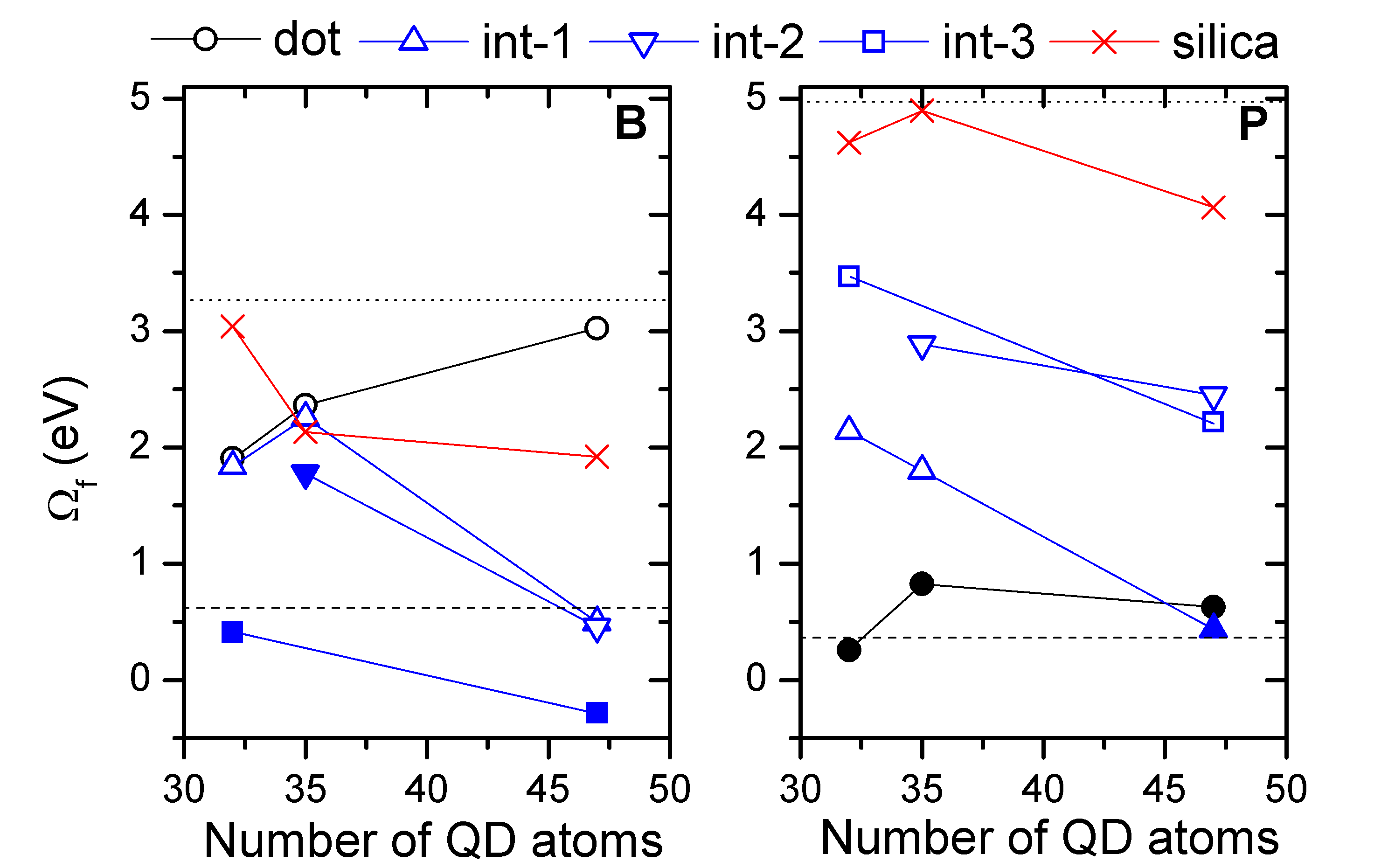}
	\caption{Formation energy $\Omega_{f}$ as a function of QD size, for different positions of B (left panel) and P (right panel) dopant atoms. Zero energy corresponds to the undoped systems. Filled symbols highlight the most stable doped configuration. The value of $\Omega_{f}$ for the impurity in bulk silicon (dashed line) and bulk silica (dotted line) is reported for comparison.}
	\label{fig:formationE-size}
\end{figure}

It is known that substitutional impurities produce a local distortion, that must be taken into account for a realistic description of doping.
In the case of free-standing hydrogenated Si-QDs, impurities located close to the QD surface are energetically more favorable than others, thanks to a larger atomic mobility that allows a reduction of the stress around the dopant atom.\cite{mavros,eom,ossicini-2006,location,ma} In this case, doping the nanostructure core region could result very difficult, even for materials commonly doped in their bulk phase.\cite{mavros}
Beside strain effects, other chemistry-governed factors, occurring at shorter scales, can determine the energetically favored site of the impurity. For example, in the case of OH-terminated or SiO$_2$-embedded QDs, the strong electronegativity of O makes P strongly repelled from the interfacial sites, while conversely attracting B.\cite{guerrajacs,carvalho1,carvalho2,xie-B}
This behavior has been observed in free-standing Si-QDs experiments,\cite{free-standing-P-1,free-standing-P-2} and suggested as the mechanism for IR absorption in B-doped free-standing Si-QDs, not observed in P-doped ones.\cite{location}

In Figure~\ref{fig:formationE-size} we report the formation energy $\Omega_{f}$ of all the considered structures and doping sites, calculated following Ref.~\cite{ossicini-2006}:
\begin{equation}
	\Omega_{f}=E_{doped}-E_{undoped}+E_{Si}-E_{dopant},
	\label{eq:E-formation}
\end{equation}
where E$_{undoped}$ and E$_{doped}$ are the total energies of the undoped and doped systems, respectively, E$_{dopant}$ is the total energy per atom in a bulk configuration of the dopant atom,\cite{dopant-bulk} and E$_{Si}$ is the total energy per atom of bulk silicon.

Consistently with the above discussion, we note in Figure~\ref{fig:formationE-size} that P and B impurities prefer to site inside QD and at the interface, respectively. Moreover, while the formation energy in P-doped systems decreases with the suboxide number, it conversely increases for B-doping.
Interestingly, we also note that for the largest considered QD, Si$_{47}$, the placement of P in the QD center is energetically similar to the int-1 case. This is because the QD core-region cannot easily accomodate the impurity-induced stress.\cite{eom} Therefore, a more energetically stable site should be found close to the interface (in order to accomodate the stress more easily), but still inside the QD (to take advantage of the P-Si bond over P-O).
The latter argument is supported by XPS measurements showing a clear B-O bond signal,\cite{xie-B} while P-Si or P-P bonds seem preferentially formed rather than P-O bonds for Si-QDs with diameters smaller than 3.5\,nm.\cite{perego,cho-P} Moreover, also PL experiments suggest B-doped Si-QDs with an intrinsic core and heavily B-doped shells,\cite{B-surface} while B-P codoped colloidal Si nanocrystals show an outer B-rich shell and an inner P-rich shell, arising due to the large difference in the segregation coefficient of B and P.\cite{deliver-16,deliver-17}

In Fig.~\ref{fig:formationE-size} we also report the $\Omega_{f}$ of doped bulk-Si (dashed line) and a-SiO$_2$ (dotted line). Clearly, the formation energy for doped a-SiO$_2$ is much higher than for doped bulk-Si, especially for P-doping, on line with recent experiments observing P-atoms segregating toward the Si-rich regions.\cite{perego} Also, secondary ion mass spectroscopy (SIMS) experiments confirmed a negligible B or P diffusion from Si-QD layers to adjacent SiO$_2$ layers.\cite{hao-B,hao-P}

\begin{figure}[t!]
	\centering
	\includegraphics[width=8.2cm]{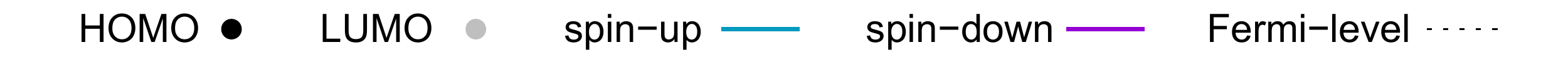}
	\includegraphics[width=8.2cm]{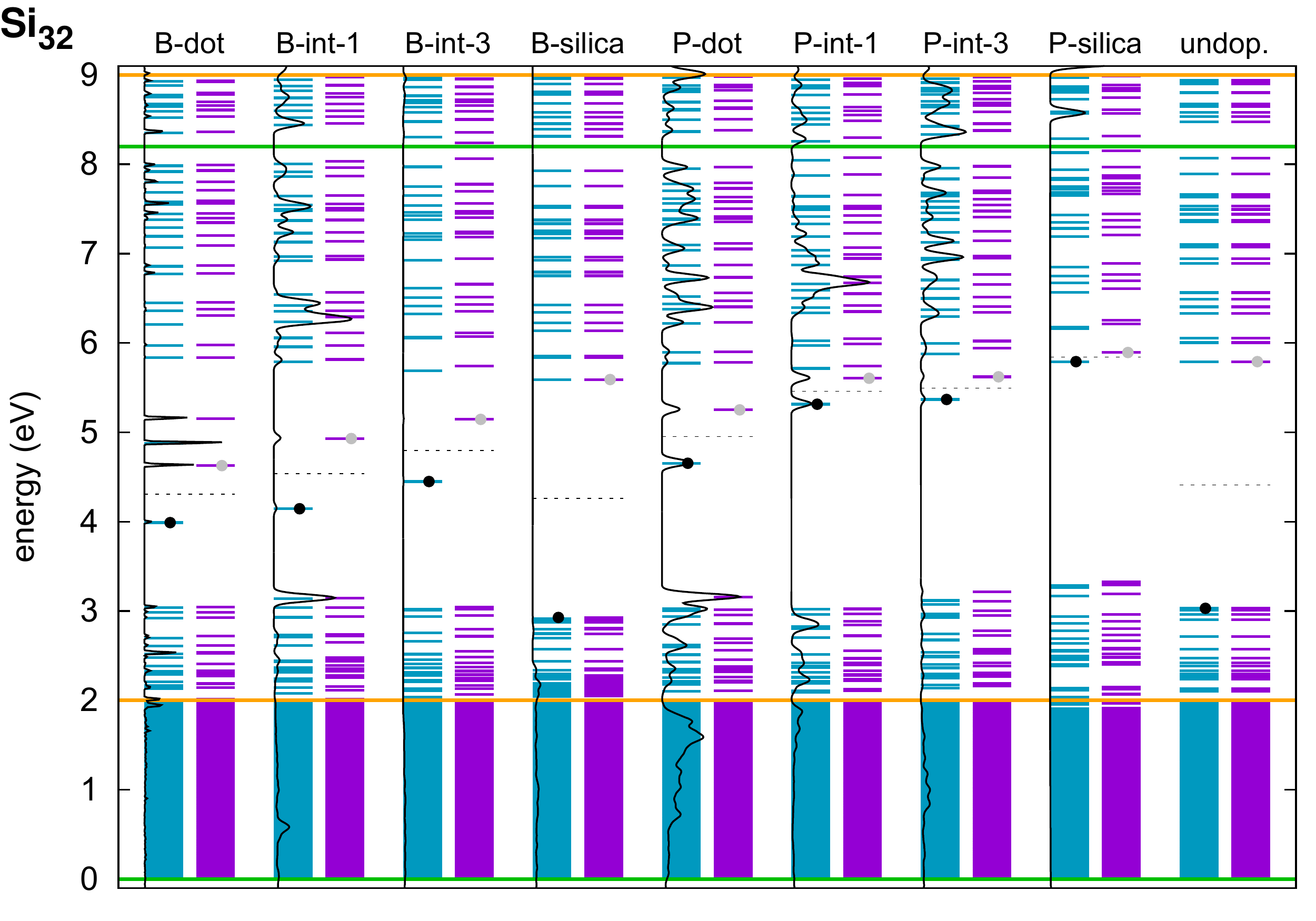}
	\includegraphics[width=8.2cm]{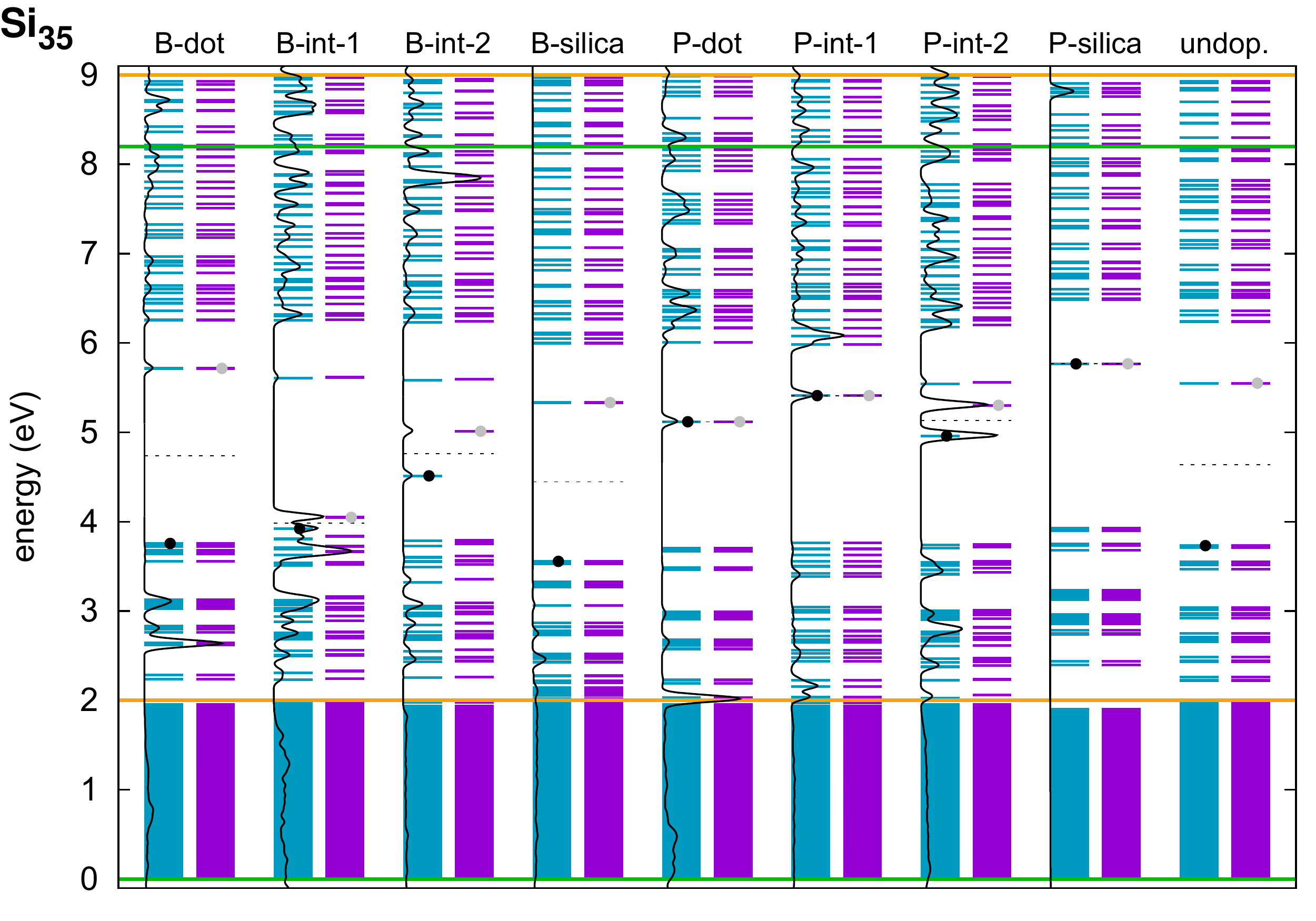}
	\includegraphics[width=8.2cm]{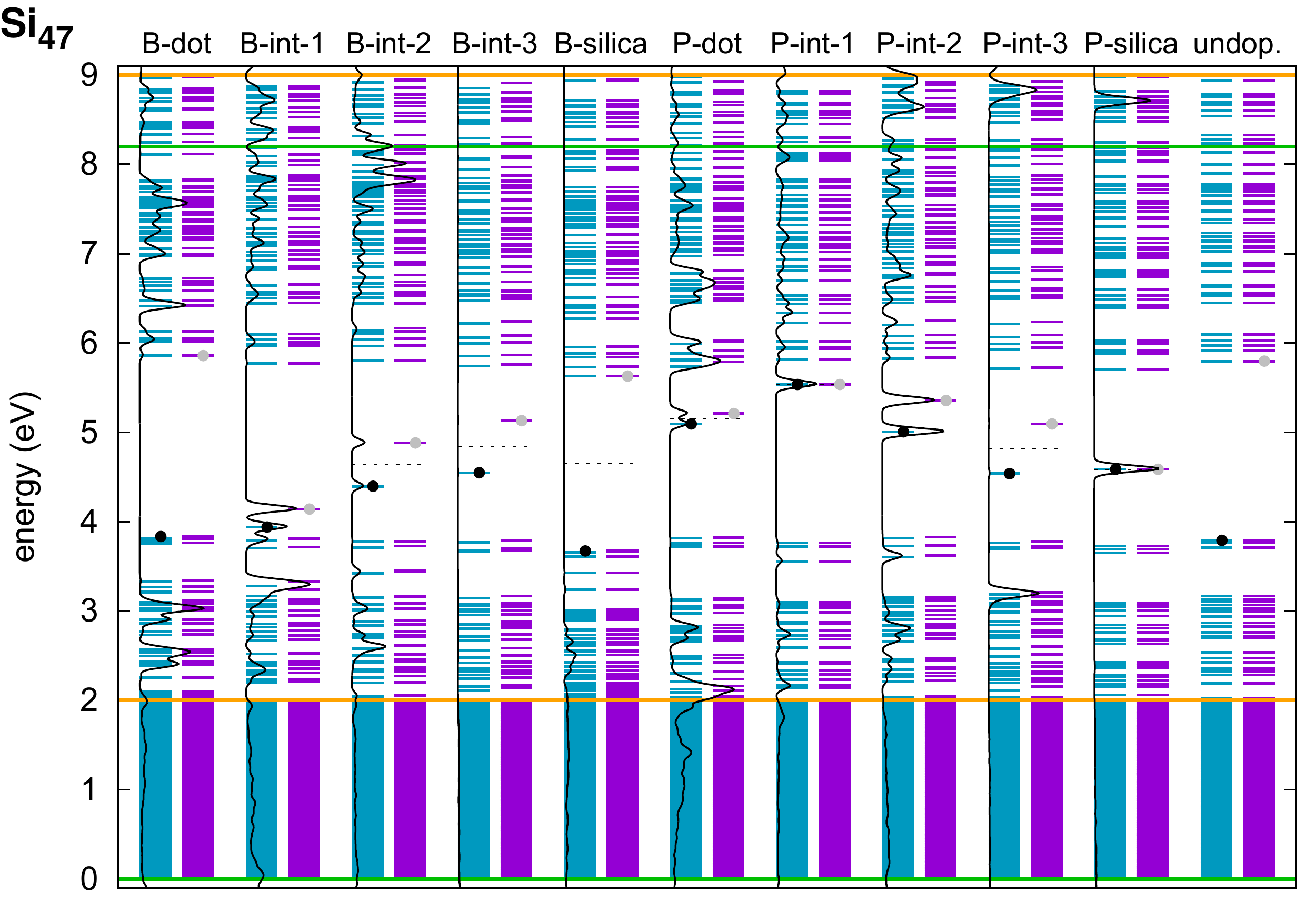}
	\caption{Spin-up and spin-down eigenvalue spectrum of all the considered systems: Si$_{32}$ (top), Si$_{35}$ (center), and Si$_{47}$ (bottom). Energies have been aligned using the states of the embedding SiO$_2$ (see Method). For each case, the PDOS of the dopant atom is also reported (black solid line). Black and grey dots mark the HOMO and LUMO states, respectively, while $E_F$ is reported by dashed line. Horizontal lines mark the uncorrected (orange) and corrected (green) band-edge of SiO$_2$ (see Method).
	}\label{fig:EIG}
\end{figure}

\subsection{Electronic properties} \label{sec.electronic_properties}

The inclusion of impurity atoms in the pristine system leads to a reduction of $E_g$ due to the appearance of mid-gap states, whose energy and localization can vary in a very complex way.\cite{mavros,mavros-54}
In our systems, doping with single group-III or group-V impurities results in an odd number of electrons, for which spin-polarized calculations must be employed. For small Si-QDs, a correlation between the energy difference of spin-down and spin-up impurity levels
and the magnitude of the Stokes shift of undoped Si-QDs has been reported, signaling structural deformation.\cite{ramos}

In Figure~\ref{fig:EIG} we report the eigenvalues of all the systems, with energies aligned using the embedding-SiO$_2$ states (in order to get the band offset, see Method; see Supplementary Information\dag for numerical data). For the doped systems we also report the PDOS of the dopant atom.

The expected acceptor behavior of B impurities -- lowering of the Fermi energy toward the valence band -- which is clearly observed in hydrogenated Si-QDs,\cite{eom} is only present in some of our embedded systems.
Instead, in most of our structures the impurity generates deep levels, not a suitable condition for enhanced current transport. Besides, the dramatic reduction of $E_g$ occurring in many cases, is an important requirement to foster the conductivity at low $V$.
In the case of P impurities we observe a clear donor behavior, as occurring in free-standing n-doped Si-QDs.\cite{mavros}

It is worth to stress that the variability of the observed response with doping conditions, among the three considered QDs, is expected due to the large QD surface-to-volume ratio.\cite{DFT2}
Nevertheless, it is possible to recognize some trends in our data. First, for QDs B-doped at the interface (the most energetically favored site for B) $E_F$ increases with the suboxide number, with correspondingly increasing HB and decreasing EB. Conversely, the interfacial P-doping produces an n-type effect, with $E_F$ slightly decreasing with the suboxide number.

The doping at SiO$_2$ sites, far from QD, produces for B impurity a minimal decrease of $E_g$ (and of $E_F$), that should lead to a conductivity similar to the undoped case. The same behaviour is found for B-doping at the center of the QD.
In the case of P impurity, $E_g$ is dramatically reduced in all the cases, while no clear trend for $E_F$ can be deduced. Unfortunately, as discussed above, any potential advantage of P-doping at SiO$_2$ sites is limited by its strongly unfavored energetics.
However, $E_g$ is reduced also in the case of P located at the QD center or at interfacial sites with low suboxide number (the most favored sites), in which case we also observe HOMO and LUMO states populated by the PDOS of the impurity atom. In this case we expect a significant enhancement of electron current, also at low $V$.

\begin{figure}[t!]
	\centering
	\includegraphics[width=8.2cm]{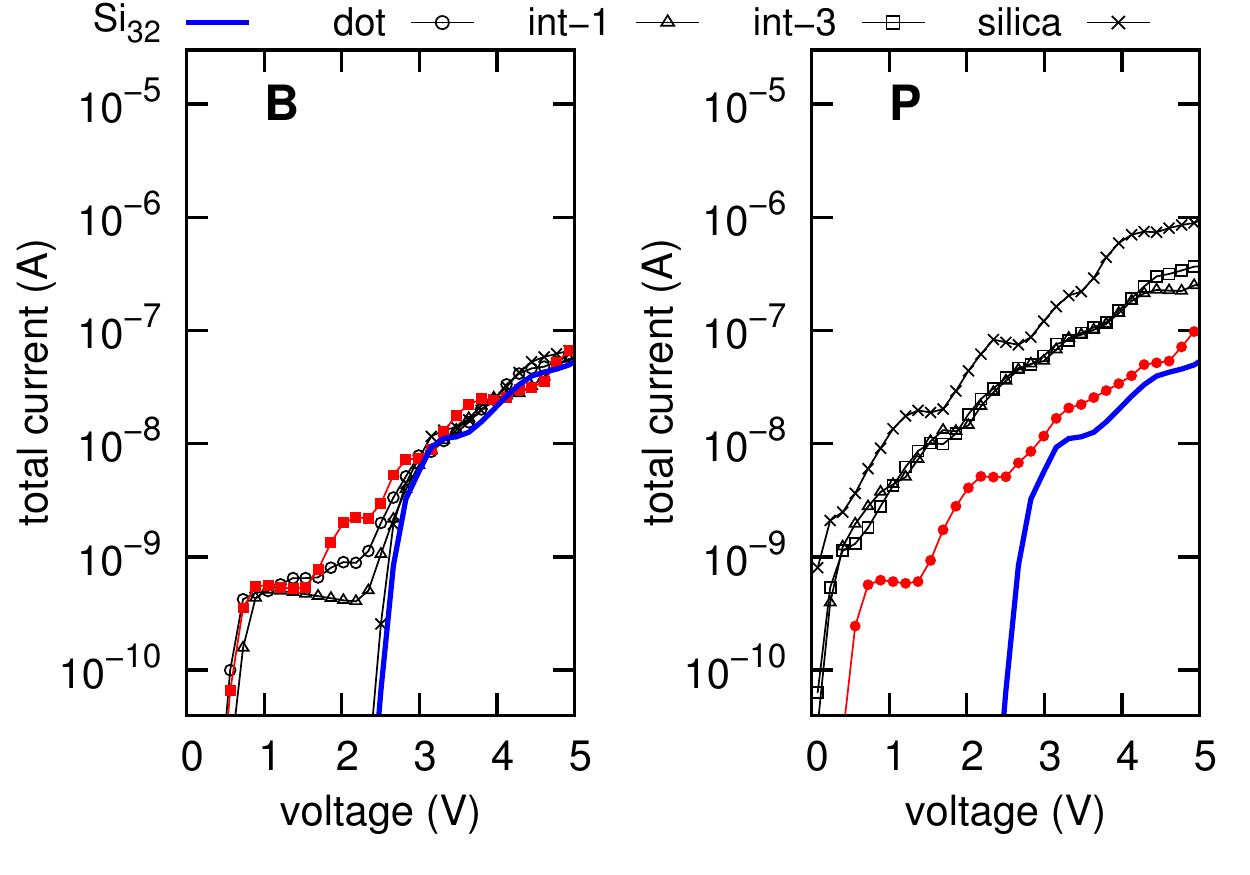}
	\includegraphics[width=8.2cm]{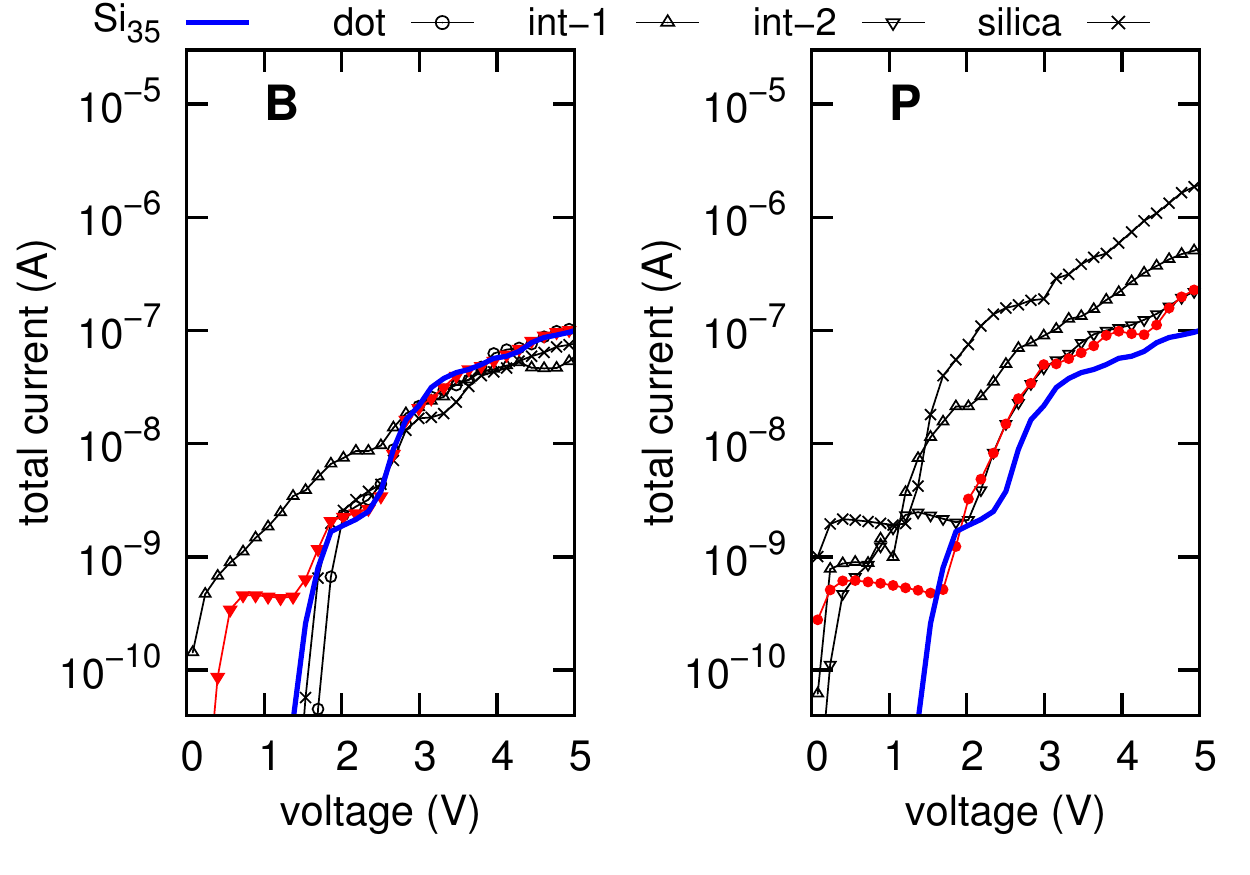}
	\includegraphics[width=8.2cm]{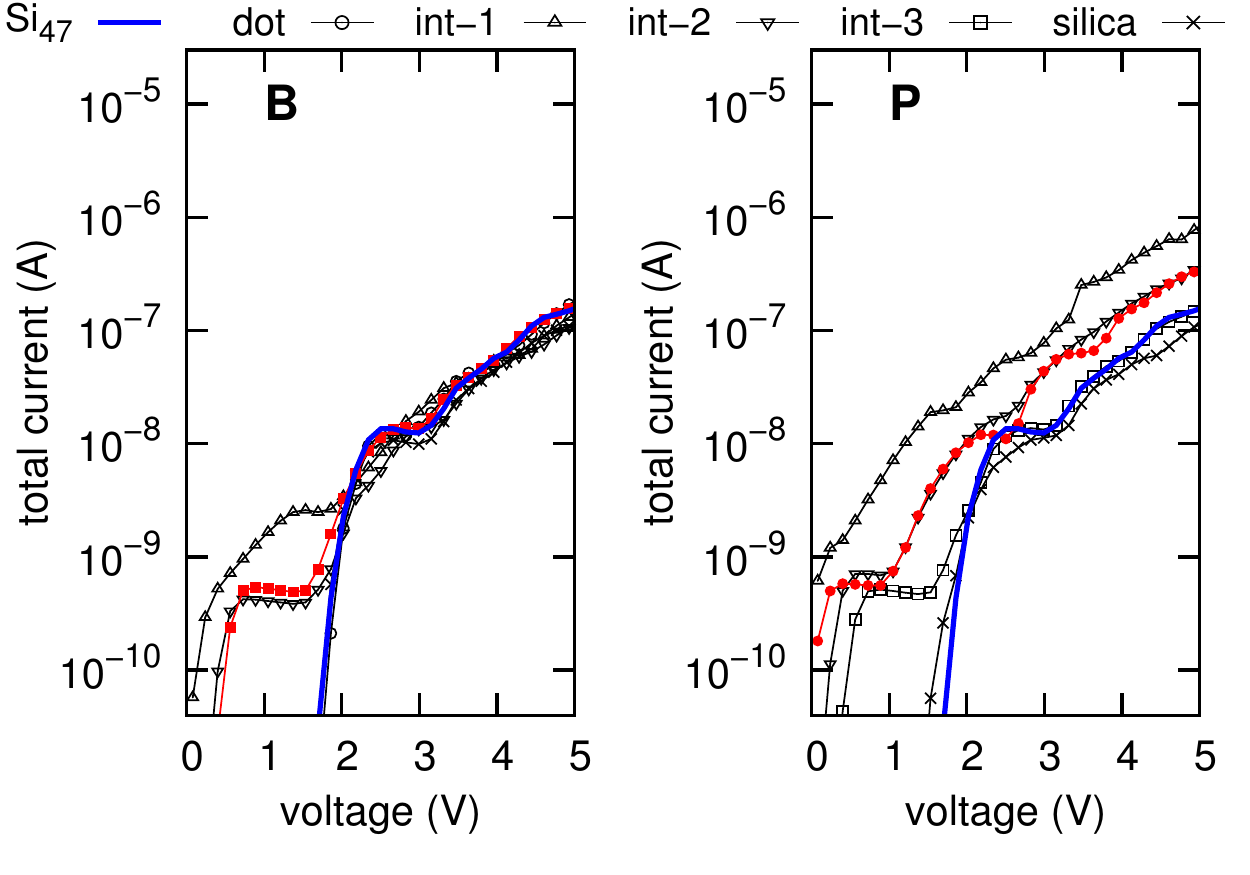}
	\caption{Calculated total current (electron + hole) as a function of the applied voltage, for the considered doped configurations (symbols) along with the undoped case (blue solid curve), for Si$_{32}$ (top), Si$_{35}$ (center), and Si$_{47}$ (bottom) QDs. Filled symbols (in red) highlight the most stable doped configuration (See also Supplementary Information\dag).
	}\label{fig:IV}
\end{figure}

\subsection{Transport properties} \label{sec.transport_properties}

In Figure~\ref{fig:IV} we show the computed $I$-$V$ characteristic of each system, with total current obtained by summing electron and hole currents (See Supplementary Information\dag for separate electron/hole $I$-$V$ curves). The results are compared with the corresponding undoped case\cite{IV-diameters} (the separate electron and hole contributions to the total current, as a function of the applied voltage, are reported in the Supplementary Information\dag).
In the results of Fig.~\ref{fig:IV} are reflected all the above-discussed effects of doping over the electronic properties: as $E_F$ approaches the conduction (valence) band, electron (hole) barrier EB (HB) becomes smaller, and the electron (hole) current is enhanced with respect to the undoped system. Instead, the threshold $V$ for triggering transport is related to $E_g$ -- typically reduced by doping -- that determines the ionization energy required to generate free carriers.

The latter aspect is well observed in B-doped structures, especially for interfacial doping (the one with the lowest formation energy) showing, with respect to the undoped case, a significant enhancement of the total current at low $V$ (due to $E_g$ decrease), while at large $V$ we observe no significant variation of the current.

Doping at SiO$_2$ sites seems practically irrelevant in B-doped structures, while it produces dramatic enhancements of the current, up to two orders of magnitude, for two of the P-doped structures, having although the largest formation energy.
Nevertheless, the more energetically-favored P-doping inside QD yields still an increase of the $I$-$V$ response in all the considered $V$-range, up to one order of magnitude, also at high $V$ for all the QDs.

\section{Conclusions}

We have studied Si-substitutional doping of Si-QDs embedded in SiO$_2$, with either B or P impurities.
Calculations reveal that B impurities tend to site at the QD/SiO$_2$ interface, especially where a large number of bonding oxygens is present. Conversely, doping inside the QD or the SiO$_2$ is unfavored, with similarly large formation energies. For P impurities, we have observed a clear trend in which the formation energy increases with the number of bonding oxygens, hence favoring the QD internal, while severely hampering interfacial and SiO$_2$ sites. Besides, given the large Si/SiO$_2$ interfacial energy, P-doping at interface Si$^{1+}$ sites may be favored over QD-core regions, especially in large QDs, thanks to an easier relaxation of the doping-induced stress at the interface. Therefore, we indicate sub-interfacial QD sites as the most energetically-favored ones for P impurities.

In any case, the presence of impurities reduces the band-gap $E_g$ of the material -- except for B-doping in the QD or in SiO$_2$ (the two least probable sites) -- leading to enhanced $I$-$V$ characteristic at low $V$. At high $V$, for the most favored impurity positions we observe a significant variation of the current, with respect to the undoped systems, only for P-doping.

Thus, with either B or P impurities one can foster hole-current at low $V$ or electron-current at low+high $V$, respectively. Such asymmetry of the response with the dopant type can be advantageous from a technological point of view, permitting the tuning of the device response in the given $V$ range.
For example, possible applications can extend from Si-based photonics,\cite{deliver-16} to next-generation photovoltaic solar cells,\cite{conibeer} among others.

\section*{Acknowledgements}
{\footnotesize R. G. acknowledges support from the ERC Advanced Grant no. 320796-MODPHYSFRICT. S. I. acknowledges support from the FI program of the Generalitat de Catalunya. A. C. acknowledges support from the ICREA academia program. J. D. P. acknowledges support from the Serra Hunter programme. The authors thankfully acknowledge the Barcelona Supercomputing Center - Centro Nacional de Supercomputaci\'{o}n, and the CINECA-ISCRA initiative.}


\begin{thebibliography}{}
\bibitem{dopant-bulk} Bulk systems of dopant atoms: for B atom we used the B$_{50}$ system, an alpha-tetragonal phase (doi:10.1103/PhysRevLett.90.026103) with IT number 134, space group P42/nnm, $a$\,=\,$b$\,=\,8.75\,\AA, $c$\,=\,5.06\,\AA, and angles of 90$^\circ$; for P atom we used the black P system, an orthorombic phase (doi:10.1103/PhysRevB.66.161202), with IT number 64, space group Cmca, $a$\,=\,3.314\,\AA, $b$\,=\,10.478\,\AA, $c$\,=\,4.376\,\AA, and angles of 90$^\circ$.
\end{thebibliography}
\end{document}